# A Systematic Review on Women's Participation in Agricultural Work and Nutritional Outcomes

**Pallavi Gupta (PhD Development Economics)**
Assistant Professor,
Sarla Anil Modi School of Economics,
Narsee Monjee Institute of Management Studies, Mumbai
pallavi.gupta@nmims.edu

## Abstract

While agriculture is recognised as vital for improving nutrition, the evidence linking women's participation to sustained nutritional gains remains inconclusive. This review synthesizes studies published between 2000 and 2024 to reflect current agricultural practices and nutritional challenges. We examine how agricultural practices and time use affect nutritional outcomes among rural women through pathways such as income generation, food preparation, and intra-household labour allocation. A structured methodology with clear inclusion and exclusion criteria was used to assess gender-sensitive and nutrition-sensitive interventions. Using narrative synthesis, the review categorizes findings around key themes and contextual factors, including socio-economic status, seasonality, and labour intensity. The results show that while increased involvement in agriculture can boost household dietary diversity and income, it also raises time burdens that affect food preparation, childcare, and self-care. Positive outcomes occur when interventions enhance women's decision-making power, income access, and use of time-saving technologies, whereas negative outcomes emerge when excessive workloads compromise energy balance and limit rest. A conceptual framework is presented to map the dual pathways linking agriculture, time use, and nutrition, capturing the roles of labour distribution, social norms, and resource access. The framework underscores the need to integrate gender equity, time efficiency, and nutritional objectives into agricultural policies. In conclusion, agricultural interventions have potential for nutritional improvement if they are carefully designed to avoid unintended negative impacts on women.

## Introduction

The complex relationship between agriculture, time use, and nutrition outcomes is a critical issue in global development, particularly in low- and middle-income countries where nutrition-related challenges persist. Undernutrition, especially maternal and child undernutrition, is a multifaceted problem that is influenced by various factors across different sectors, including agriculture, healthcare, education, and income. While agricultural development has long been seen as a pathway to improved food security and nutritional outcomes, evidence of its direct impact on nutrition has been mixed. Numerous systematic reviews have failed to find consistent, clear evidence linking agricultural interventions with significant nutritional improvements, especially in rural areas where agriculture plays a dominant role in livelihoods (Webb and Kennedy, 2014).

The need for effective interventions to tackle the underlying causes of undernutrition has led to an increasing recognition of the importance of agriculture in shaping nutritional outcomes. The development of agriculture-based policies that are sensitive to nutritional needs is crucial, but the pathways through which agriculture affects nutrition are still not fully understood. Bhutta et al. (2013) emphasise the necessity of widening the scope of development policies to address the deeper, structural causes of undernutrition, and agriculture remains a key sector in improving the nutritional status of populations, especially women and children. However, despite the potential for agriculture to impact nutrition positively, the evidence base remains weak due to a lack of high-quality research and evaluations that link agricultural interventions with nutritional outcomes (Girard et al., 2012; Ruel and Alderman, 2013).

The existing literature has largely focused on the macroeconomic effects of agricultural interventions, but it has overlooked the complex dynamics at the household and individual levels. Women, as primary managers of household food security, play a central role in leveraging agricultural improvements to enhance nutrition. Ruel and Alderman (2013) note that women's role in agriculture is fundamental to translating agricultural interventions into nutritional benefits. However, there is limited research on how agricultural interventions affect women's time use, their control over resources, and their ability to influence household nutritional outcomes. This gap highlights the importance of understanding the role of time use in the agriculture-nutrition nexus, particularly in the context of women's labour and how it intersects with the household's nutritional status.

Our review seeks to address these gaps by investigating the gendered dimensions of changing time use in agricultural contexts and the subsequent impact on nutritional outcomes. We aim to explore the complex relationship between agricultural practices and time use, and how these, in turn, influence nutrition. Specifically, this review focuses on two main pathways: the first connects agricultural practices and interventions with time use, and the second links time use with nutrition outcomes. The review synthesises both quantitative and qualitative evidence using a narrative synthesis approach to organise and discuss key findings across various studies. This methodology is particularly useful in organising diverse evidence and providing insights into the complex factors that mediate the relationship between agriculture, time use, and nutrition.

Agricultural interventions, while potentially beneficial, often come with increased time burdens, especially for women who are already tasked with multiple roles within the household. While women's labour in agriculture has traditionally been undervalued, there is growing recognition that their involvement in agricultural production, both for subsistence and income, directly impacts the household's nutrition and food security. Studies have shown that agricultural interventions frequently increase time spent in farming activities, which can reduce the time available for other crucial tasks, such as food preparation, child care, and household maintenance. This time burden can potentially affect women's nutritional status and well-being, leading to unintended negative consequences.

However, the link between increasing time burdens and negative nutritional outcomes is not straightforward. Some studies suggest that time constraints can be mitigated by income gains from agricultural interventions. As households gain more income, they may be able to purchase more food or even hire labour to assist with household and agricultural tasks. This income effect can potentially offset the negative impacts of rising time burdens, leading to improved dietary diversity and nutritional outcomes. However, this outcome depends on the type of intervention, the socioeconomic status of the household, and the specific dynamics within the household, especially regarding the allocation of labour and resources (Kadiyala et al., 2014).

The evidence suggests that agricultural interventions tend to increase agricultural time commitments, but the impact on nutrition is more complex. In some cases, the increased time spent in agriculture may lead to a shift away from home-grown food production and preparation, with households relying more on purchased foods. While this may increase calorie intake, it does not necessarily lead to improved nutritional status. The type and quality of foods consumed, especially micronutrient-rich foods, play a critical role in determining nutritional outcomes. Furthermore, time use patterns within households often reflect broader gender dynamics, with women shouldering a disproportionate amount of unpaid labour. This gendered division of labour complicates the impact of agricultural interventions on nutrition, as women's increased participation in agricultural work may reduce their time for other activities that directly affect household nutrition.

This research is significant for several reasons. First, it provides a comprehensive summary of evidence on the time burdens associated with agricultural interventions, particularly for women. Second, it offers new insights into the complex relationship between time use and nutritional outcomes, emphasising that the impact of agricultural interventions on nutrition is not always straightforward. Third, it provides recommendations for future research to address the gaps in the current evidence base. Specifically, future research should focus on improving time use data, incorporating multiple indicators of nutrition, and expanding the scope of studies to include other household members, particularly men and children. Additionally, more research is needed to explore the role of time-saving technologies and the intra-household allocation of labour in mitigating the impact of rising time burdens.

**Methodology and Search Strategy**

To compile this literature review, an extensive search was conducted using academic databases, including Google Scholar, JSTOR, ScienceDirect, and PubMed, as well as grey literature and reports from government agencies and NGOs. The primary search terms included "women's time use in agriculture," "agriculture and nutrition," "women's empowerment and nutrition," "gender and food security in rural India," and "agricultural interventions and women's health." The search was further refined by focusing on studies conducted in rural India.

**Inclusion Criteria**

To ensure the relevance and quality of studies, the following four inclusion criteria were applied:

a. **Geographic Focus**: Studies focusing on rural India or similar agrarian economies.
b. **Time Frame**: Studies published between 2000 and 2024 to ensure relevance to current agricultural practices and nutritional challenges.
c. **Study Type**: Peer-reviewed articles, government reports, and reputable institutional publications that explore the intersection of agriculture, women's time use, and nutrition.
d. **Themes**: Studies must address one or more of the following themes: (a) women's agricultural labour, (b) time-use patterns in agriculture, (c) nutritional outcomes, (d) women's empowerment and decision-making in agriculture, and (e) the impact of agricultural interventions on nutrition.

**Exclusion Criteria**

The following studies were excluded from the review:

a. Studies that are not focused on rural or agricultural contexts.
b. Studies that did not provide clear connections between agriculture, time use, and nutrition, or those that focused solely on urban settings.
c. Studies that primarily dealt with non-empirical data, opinion pieces, or those that were not peer-reviewed.

d. Articles that failed to address gender-specific aspects of agricultural work or lacked a clear focus on women's roles in agriculture.

 Key Findings

**Women's Role in Agriculture and Time Use Patterns**

Women's involvement in agriculture has been a subject of growing academic interest, particularly in light of the feminisation of agriculture. Studies like Gunewardena (2010) and Roberts (1996) indicate that women's participation in agriculture has significantly increased, especially in labour-intensive tasks such as sowing, harvesting, and post-harvest processing. In rural India, the shift to cash crop production has placed a greater burden on women, who increasingly participate in commercial agriculture. This has led to higher time burdens, as women are tasked with managing both agricultural work and domestic duties.

Mishra and Mishra (2012) and Zaman (1995) found that women often spend more time than men on agricultural activities, particularly during peak seasons. Women's time is further constrained by domestic work, such as cooking, cleaning, and child care. These dual responsibilities often lead to time poverty, reducing women's ability to prioritise food preparation and child care, which are crucial for good nutrition.

**Agricultural Interventions and Time Burdens**

While agricultural interventions aim to improve food security and nutrition, many studies have highlighted the unintended consequences of increased time burdens on women. Studies by Kumar (1994) and Paolisso et al. (2002) found that interventions such as the adoption of hybrid crops or cash crop programs increase women's time spent on agricultural labour, often at the expense of domestic responsibilities. Although agricultural income may increase as a result of these interventions, the time spent on agricultural tasks reduces the time available for food preparation, which can negatively affect dietary diversity and nutritional outcomes.

However, studies like Quisumbing et al. (2013) and Riley and Krogman (1993) suggest that agricultural interventions can have positive nutritional outcomes, provided they also address the income aspects. When households have more income, they may be able to purchase more diverse and nutritious foods, which could mitigate the time burdens associated with agriculture.

**Women's Empowerment and Nutritional Outcomes**

Women's empowerment has been shown to have significant positive effects on nutrition, particularly in households where women have decision-making power over income and food resources. Malapit and Quisumbing (2015) and Sraboni et al. (2014) found that women's control over income and agricultural resources led to better household dietary diversity and improved child nutrition. In India, studies by Gupta et al. (2017) highlight the positive association between the Women's Empowerment in Agriculture Index (WEAI) and household nutritional outcomes, with empowered women being more likely to ensure a diverse and nutritious diet for their families.

However, despite the positive effects of women's empowerment, challenges persist in ensuring that these benefits are translated into improved nutrition. Empowerment alone does not guarantee that women will have the time or resources to prepare nutritious meals, particularly in the face of time poverty.

**Socio-Economic Status and Household Composition**

Socioeconomic status plays a crucial role in determining nutritional outcomes, as wealthier households are better able to manage the time burdens associated with agricultural work. Studies by Kadiyala et al. (2014) and Zaman (1995) indicate that wealthier households can hire labour for both agricultural and domestic tasks, reducing the time burdens on women and allowing them to focus on other aspects of household management, including food preparation. Conversely, poorer households, particularly those without access to hired labour, are more likely to face increased time constraints, which may lead to lower nutritional outcomes.

Household composition also plays a significant role in time use and nutritional outcomes. In households with more members, there is often a redistribution of tasks, with younger members or other women taking on some of the domestic work. However, this redistribution is influenced by gender and age, as women and children are more likely to bear the brunt of the additional work, as noted by Seymour and Floro (2016).

**Seasonality and Nutritional Impacts**

Seasonality is a critical factor that influences both time use and nutrition. Agricultural work is highly seasonal, with peak labour demands during planting and harvest periods. Studies by Wodon and Beegle (2006) and Kumar (1994) show that women's time burdens are particularly high during these periods, which can reduce the time available for food preparation and child care. This seasonal variation can lead to fluctuations in household nutritional outcomes, with periods of increased labour demand corresponding to reduced food security and dietary diversity.

**The Role of Socioeconomic Factors**

Socioeconomic status plays a crucial role in mediating the effects of agricultural work on nutrition. Wealthier households can cope with increased time burdens by hiring external labour for both agricultural and domestic tasks, which allows them to allocate more time to food preparation and child care. In contrast, poor households, especially those with fewer resources, are less able to afford such labour and thus bear the full brunt of the increased work burden, often nutritional status.

**Conceptual Framework**

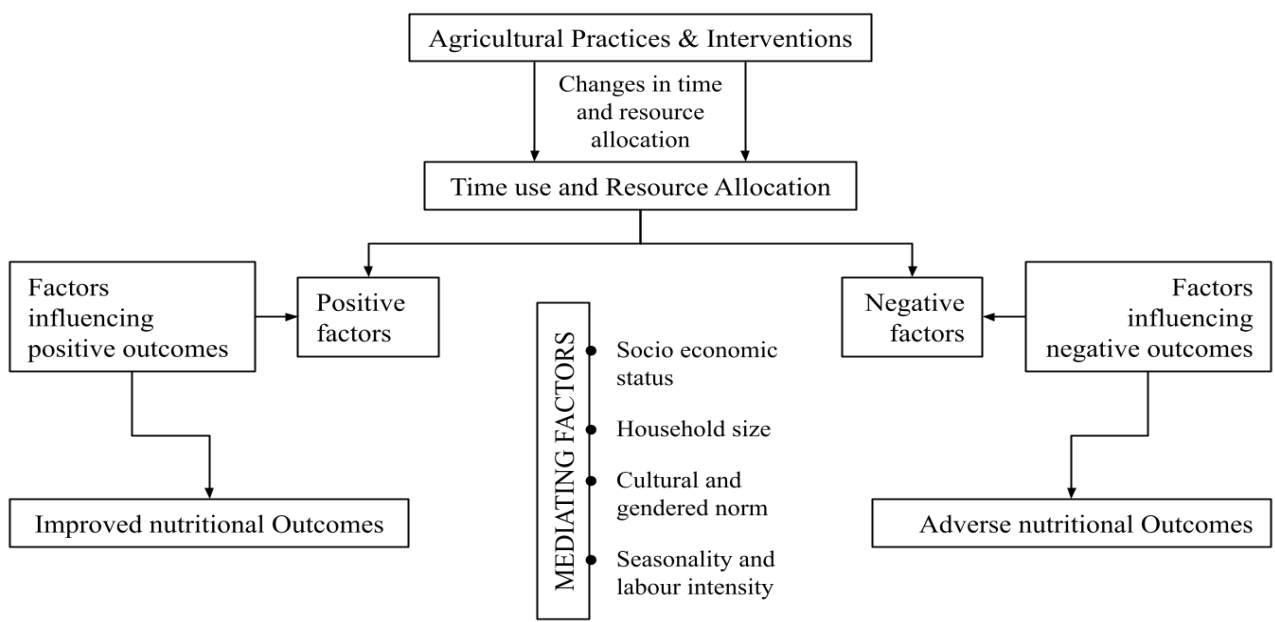

AGRICULTURE - TIME USE - NUTRITION NEXUS

The conceptual framework given above visually showcases the complexity of the agriculture-nutrition link, emphasising both enabling and limiting factors. There are three key elements of this framework: **Dual Nature of Pathways**: Factors like women's empowerment, market access, or income can lead to positive outcomes, but overburdened women or time constraints may push outcomes toward the negative side. 2. **Role of Mediating Factors**: Variables like socio-economic status or cultural norms modulate the direction of the outcome, making them critical touchpoints. 3. **Feedback Loops**: Negative nutritional outcomes, like undernutrition or poor health, can feed into the system, reducing labour efficiency or perpetuating poverty cycles.

**Discussion**

**Trade-Offs Between Economic Gains and Time Poverty**

One of the central dilemmas in women's agricultural labour is the balance between economic empowerment and time poverty. While income from agricultural work can enhance food security, increased labour burdens often lead to reduced time for self-care, food preparation, and childcare, negatively impacting nutritional outcomes (Johnston et al., 2018). Studies indicate that when women's agricultural participation increases without complementary interventions, such as access to childcare, labour-saving technologies, or equitable household labour division, negative nutritional consequences may arise, particularly in resource-poor settings (Sraboni et al., 2014).

**Contextual and Structural Factors Shaping Outcomes**

The relationship between agricultural work and nutrition is not uniform across all regions or socioeconomic groups. The extent to which women benefit from agricultural labour depends on factors such as land ownership, access to credit, control over decision-making, and the structure of local food markets (Doss, 2018). In some cases, agricultural intensification disproportionately increases women's workloads without yielding substantial nutritional benefits due to persistent gender inequities in resource access (Meinzen-Dick et al., 2019). Thus, a more intersectional approach—considering caste, class, and regional disparities—would provide a more comprehensive understanding of these dynamics.

**Intra-Household Bargaining and Nutritional Outcomes**

A deeper examination of intra-household bargaining dynamics is crucial in understanding how increased agricultural incomes translate into better nutritional outcomes. Research suggests that women's control over agricultural earnings is more strongly associated with improved dietary diversity and child nutrition compared to male-controlled income (Malapit et al., 2015). However, social norms and gendered power imbalances often limit women's decision-making power, preventing them from allocating resources optimally for household nutrition (Quisumbing et al., 2021). Future research should focus on how policies and interventions can shift these power dynamics to enhance nutritional benefits.

**Policy Gaps and the Role of Nutrition-Sensitive Agriculture**

While gender-sensitive agricultural policies have gained traction, many interventions focus primarily on economic participation rather than addressing time-use burdens or dietary outcomes. There is a need for policies that integrate gender considerations into agricultural extension programs, promote labour-saving technologies (e.g., mechanized farming tools), and improve social infrastructure such as childcare and community kitchens (Ruel et al., 2018). Furthermore, linking agricultural programs with nutrition education initiatives can ensure that increased agricultural productivity translates into healthier food choices and improved nutritional outcomes.

## Conclusion

This synthesis reveals a complex relationship between agriculture, time use, and nutrition outcomes in rural India. While increased participation in agricultural work can enhance income and food access, it also places significant time burdens on women, which can adversely affect their health and the nutritional outcomes of their families. Women's empowerment plays a critical role in mitigating the negative effects of time burdens, as it enables women to make informed decisions about food, health, and resource allocation. However, the impacts of agricultural interventions are not uniform and depend on a range of factors, including household resources, the type of agricultural activities involved, and the seasonality of work.

We also bring out a better understanding of the intricate relationship between agriculture, time use, and nutrition outcomes, highlighting lessons from existing studies. A key focus is whether shifts in time burdens, particularly for women, can explain the disconnect between agricultural productivity and nutritional improvements. While the findings affirm the significant role women play in agriculture, they also reveal that agricultural interventions often increase the time burdens on participants. However, the relationship between these increased time burdens and nutrition outcomes is complex and varies across different contexts. Increased time demands may sometimes lead to unintended negative consequences, but this is not universally the case. The study, therefore, provides valuable insights and a road map for future research into gender-sensitive and nutrition-sensitive agricultural interventions.

The review of evidence underscores several critical areas for advancing research in this domain. First, there is a clear need to enhance the quality of time-use data. Current data collection often overlooks simultaneous activities and lacks detailed measures of work intensity, which are crucial for capturing the full scope of labour contributions and their implications. Integrating time-use data with socio-economic and activity-specific indicators could offer a more nuanced understanding of how agricultural interventions impact different groups, particularly marginalised populations.

Second, future research would benefit from employing a broader range of indicators to measure nutrition outcomes. Beyond calorie intake, metrics such as dietary diversity and anthropometric measures should be incorporated. Third, the review highlights the need to expand the focus beyond women's time use to include other household members, such as men and children. While women's contributions to agriculture and domestic labour are central to this discussion, understanding how time use is distributed across the household can illuminate intra-household labour dynamics. For example, analysing how time-saving technologies influence labour allocation within households could reveal power relations and potential shifts in responsibilities among family members, especially between younger and older women.

Finally, trade-offs in time use require a more nuanced exploration. For instance, if an agricultural intervention improves nutritional outcomes while reducing women's leisure time, its overall evaluation depends on whether the benefits outweigh the costs. Future research should strive to balance nutritional outcomes with other dimensions of well-being, such as mental health, social participation, and rest.